\documentclass[ reprint, aps,]{revtex4-2}
\usepackage{amssymb}
\usepackage{graphicx}
\usepackage{dcolumn}
\usepackage{bm}

\begin{document}

\title{Preheating constraints in $\alpha$-attractor inflation and
Gravitational Waves production}
\author{K. El Bourakadi}
\email{k.elbourakadi@yahoo.com}
\author{Z. Sakhi}
\email{zb.sakhi@gmail.com}
\author{M. Bennai}
\email{mdbennai@yahoo.fr}
\date{\today }

\begin{abstract}
We propose a scenario where preheating occurs for a specific duration that
is parametrized by an e-folds number $N_{pre}$, our results suggest a direct
correlation between the preheating duration and the density of gravitational
waves (GWs) produced during this phase. Moreover, we investigate the
consequences of the inflationary parameters on the $\alpha$-attractor E
model in the small $\alpha$ limits. In this framework, we perform
investigations on the preheating parameters involving the number of e-folds $%
N_{pre}$, and the temperature of reheating $T_{re}$, then we show that the
parameter $n$ associated with the E model of $\alpha$-attractor inflation
has a negligible effect on the preheating duration, and we demonstrate that
gravitational wave generation during preheating satisfies the restrictions
from Planck's recent data.
\end{abstract}

\preprint{APS/123-QED}

% Force line breaks with \\
%\thanks{Gravitational Waves from Preheation in $\alpha$-attractor inflation}%

\affiliation{ Physics and Quantum Technology Team, LPMC, Ben M’sik Faculty
of Sciences, Casablanca Hassan II University, Casablanca, Morocco } 
\altaffiliation[Also at ]{LPHE-MS Laboratory Department of Physics,
Faculty of Science, Mohammed V University in Rabat, Rabat, Morocco} 

\affiliation{ LPHE-MS Laboratory Department of Physics, Faculty of Science, Mohammed V University in Rabat, Rabat,
Morocco }%
\altaffiliation[Also at ]{LPHE-MS Laboratory Department of Physics, Faculty of Science, Mohammed V University in Rabat, Rabat,
Morocco}

%\affiliation{ Physics and Quantum Technology Team, LPMC, Ben M’sik Faculty of Sciences, Casablanca Hassan II University, Casablanca, Morocco \textbackslash\textbackslash
%}%
%\altaffiliation[Also at ]{LPHE-MS Laboratory Department of Physics, Faculty of Science, Mohammed V University in Rabat, Rabat,
%Morocco} %\collaboration{MUSO Collaboration}%\noaffiliation

%\affiliation{
%\affiliation{
%\affiliation{%




%\tableofcontents
\maketitle
\section{\label{sec:1} Introduction}

Among the most exciting theories that explain the origins of the Universe is
the inflation scenario \cite{A1,A2,A3}, this modern theory described through
various models gives a satisfying solution to the standard Big Bang
problems. Accurate measurements of the cosmic microwave background (CMB) and
the large-scale structure (LSS) of galaxies provided helpful constraints on
inflation models \cite{A4}, taking into consideration the recent
observations of the $\alpha $-attractor-type models, which were suggested in
a unified approach combining the Higgs and Starobinsky inflation \cite{A5,A6}
show good compatibility with data \cite{A21,A22}; Reheating is a phenomenon
that was proposed as a technique for achieving the hot big bang Universe. At
reheating, the energy of the inflaton field is transformed to thermal
radiation through processes that may involve particle production and
non-equilibrium events, many investigations have been conducted on reheating
mechanisms \cite{A7,A8,A9,A10,C8}. However, recent studies have examined new
formalism to constrain the reheating temperature and e-folding number \cite%
{A11,A12,A13,B4}. Preheating on the other hand is a phase that occurs at the
end of inflation when the periodical oscillations of the inflaton lead to
the parametric resonance, which results in an abundant production of light
particles whose mode momenta are in the resonant bands \cite{A14,A15,A16,A17}%
. Preheating can accelerate the thermalization of our Universe since the
inflaton energy can be transferred rapidly into matter. Due to the fact that
only modes with their momenta in the resonant bands grow exponentially, the
matter distribution has large time-dependent inhomogeneities in the position
space. Thus, the matter produced during preheating has time-dependent
quadruple moments and for this reason, it could be an effective source of
gravitational waves (GWs) \cite{A18,A19,A20,D1}. The detection of GWs
produced during preheating can contribute to the investigation of inflation
and study of the reheating process.

One of the biggest difficulties confronting observational cosmology in the
coming decade is the detection of primordial tensor fluctuations, both in
the form of CMB B-mode polarization on large angular scales and a spectrum
of relic gravitational waves. Because of the direct relationship of
primordial GWs to the inflaton potential in the case of the minimum inflaton
coupling to gravity, the GW primordial power spectrum at large scales
provides vital information on the nature of the inflaton field \cite{A23}.
Equally significant is the fact that the GWs spectrum $\Omega _{GW}$ may act
as an important probe to physical processes happening after inflation. In
this paper, we investigate the preheating epoch from the $\alpha $-attractor
inflation focusing on the E-model class of $\alpha $-attractors. Our
approach to constrain preheating and the GWs produced at this stage differs
from the previous studies since we consider a preheating epoch characterized
by an e-folding number $N_{pre},$ in this way we will extend the formalism
developed in Ref. \cite{C7} where they include the nonperturbative effects
in the reheating scenario, in this way inflation is ended by an initial
nonperturbative stage followed by a perturbative one. In our work, the
reheating duration is determined through the final reheating temperature,
this leads to two phases scenarios where preheating and reheating occur for
specific durations. On another hand, we prove that the gravitational wave
spectrum $\Omega _{GW}$ is sensitive to the e-folding number $N_{pre}$,
which makes it possible to set a correlation of gravitational wave energy
density with the spectral index $n_{s}$ detected by the CMB experiments in
order to satisfy the constraints from Planck's data.

The paper is organized as follows. In Sect. \ref{sec:2} we compute the
relations between inflationary and preheating parameters in the $\alpha $%
-attractor E-model and discuss the constraints from preheating on the
inflationary parameters, we also briefly describe our proposed two phases
scenario following inflation. In Sect. \ref{sec:3} we convert the GWs\
spectra into physical variables and describe gravitational waves from
Planck's measurements perspective. In Sect. \ref{sec:4} we present our
conclusions.

\section{\label{sec:2}Preheating constraints in $\protect\alpha $-attractor
E-model}

\subsection{\label{sec:21}$\protect\alpha $-attractor inflation}

In this work we focus on a class of single-field inflation models of the $%
\alpha $-attractors, This class of inflation models can be generated by
spontaneously breaking the conformal symmetry \cite{C1}, we consider the
E-model potential in Eq. (\ref{P1}), knowing that after the slow-roll
approximation during inflation, the Klein-Gordon equation obeys $3H\dot{\phi}%
+V=0,$ while the Friedmann equation is given as$\ H^{2}=V(\phi )/3M_{p}^{2},$%
. 
\begin{equation}
V=V_{0}\left( 1-e^{-\sqrt{\frac{2\kappa ^{2}}{3\alpha }}\phi }\right) ^{2n},
\label{P1}
\end{equation}%
here, $V_{0}$, $n,$ and $\alpha $ are free parameters$.$

In the standard inflation the slow-roll parameters are introduced as $%
\epsilon =(1/2)(M_{p}V^{\prime }/V)^{2}$\ and $\eta =$\ $M_{p}^{2}V^{\prime
\prime }/V,$ which could be related to the spectral index and
tensor-to-scalar ratio as 
\begin{eqnarray}
n_{s} &=&1-6\epsilon +2\eta , \\
r &=&16\epsilon ,
\end{eqnarray}%
the energy density during inflation is written as $\rho =(1+\epsilon /3)V.$\
At the end of this epoch the energy density of the Universe can be given as $%
\rho _{end}=(4/3)V_{end}$\ considering that $\epsilon =1.$

To begin, we look for the scalar spectral index and tensor-to-scalar ratio
in the small $\alpha $ limits. In this case, we may rewrite the E-model
potential Eq. (\ref{P1}) as follows:\ 
\begin{equation}
V=V_{0}\left[ 1-2n\exp \left( -\sqrt{\frac{2\kappa ^{2}}{3\alpha }}\phi
\right) \right] ,  \label{P2}
\end{equation}%
one can compute the number of e-foldings between the horizon exit of the
pivot scale and the end of inflation as

\begin{eqnarray}
N &=&\frac{1}{M_{p}^{2}}\int_{\phi _{end}}^{\phi _{k}}\frac{V}{V^{\prime }}%
d\phi  \\
&=&\frac{1}{M_{p}^{2}}\sqrt{\frac{3\alpha }{2\kappa ^{2}}}\left[ \frac{1}{2n}%
\sqrt{\frac{3\alpha }{2\kappa ^{2}}}e^{\sqrt{\frac{2\kappa ^{2}}{3\alpha }}%
\phi }-\phi \right] _{\phi _{end}}^{\phi _{k}},  \nonumber
\end{eqnarray}%
using the approximations $\phi _{k}>>$\ $\phi _{end},$ and $M_{p}e^{\sqrt{%
\frac{2\kappa ^{2}}{3\alpha }}\phi }>>\phi $, the field at the horizon
crossing is then given as

\begin{equation}
\phi _{k}=\sqrt{\frac{3\alpha }{2\kappa ^{2}}}\ln \left( \frac{4n}{3\alpha }%
N\right) .
\end{equation}%
The above equations considering the small limit of $\alpha $ predicts

\begin{equation}
n_{s}=1-\frac{2}{N},~~r=\frac{12\alpha }{N^{2}}.  \label{Obs}
\end{equation}

\begin{table}[h]
\caption{Testing the compatibility of the Tensor-to-scalar ratio parameter
for $0.05 \leq \protect\alpha \leq 0.8$ in the case of $N=50$ and $N=60$
inflationary $e$-folds. }
\label{table:1}
%The best place to locate the table environment is directly after its first reference in text
\begin{ruledtabular}
\begin{tabular}{lcdr}
\textrm{{tensor-to-scalar ratio}}&
\textrm{$\alpha=0.8$}&
\multicolumn{1}{c}{\textrm{$\alpha=0.2$}}&
\textrm{$\alpha=0.05$}\\
\colrule
r  (N= 50) & 0.0038 & 0.0009 & 0.0002 \\
r  (N= 60) & 0.0026 & 0.0066 & 0.0016 \\
\end{tabular}
\end{ruledtabular}
\end{table}
The expression of the scalar field at the end of inflation $\phi _{end}$ is
obtained by solving the equation\ $\epsilon (\phi _{end})=1,$ which gives 
\begin{equation}
\phi _{end}=\sqrt{\frac{3\alpha }{8\kappa ^{2}}}\ln \left( \frac{4n^{2}}{%
3\alpha }\right) .  \label{end}
\end{equation}
According \cite{A11,C2} we know that $r=2H_{k}^{2}/\pi ^{2}M_{p}^{2}A_{s}$,
using the expressions above we can derive $H_{k}$ as a function of $n_{s}$
and $A_{s}$

\begin{equation}
H_{k}=\pi M_{p}\sqrt{\frac{3}{2}\alpha A_{s}}\left( 1-n_{s}\right) ,
\label{Pe}
\end{equation}%
we can determine $V_{end}=V(\phi _{end})$ using the expression of the scalar
field at the end of inflation given by Eq. (\ref{end}), we should note here
that $V_{0}$ is chosen to take the value $V_{0}^{1/4}=10^{15}GeV$ \cite{C8}.
From Table (\ref{table:1}) the inflationary parameter $r$ presents good
consistency for small $\alpha $, and the spectral index $n_{s}$ in Eq. (\ref%
{Obs}) can best fit observations for $N=60.$ However, the energy density at
the end of inflation\ gives additional constraints on the $\alpha $
parameter in order to have a valid energy $V_{end}>0,$ according to Eq. (\ref%
{end}) the parameter must be bounded as $\alpha \lesssim 0.3.$

\subsection{\label{sec:22}Preheating constraints}

At the early stages of the Universe's evolution, the process of preheating
occurs. One of the important parameters which take specific values in each
step throughout the evolution of the Universe is the equation of state (EoS)
parameter, which is defined as the ratio of the pressure and energy density
of the Universe $\omega =p/\rho $. The choice of the right EoS value after
inflation is essential, especially during preheating. Since the Universe
cools as it expands, there must be a period that follows inflation that
prepares thermally for the reheating step, that we call preheating. To
extract information about preheating we need to consider the phase between
the time observable CMB scales crossed the horizon to the present time,
which can be described by \cite{B4} 
\begin{equation}
N_{pre}=\left[ 61.6-\frac{1}{4}\ln \left( \frac{V_{end}}{\gamma H_{k}^{4}}%
\right) -N\right] -\frac{1-3\omega }{4}N_{re},  \label{pr}
\end{equation}%
the reheating duration $N_{re}$ can be obtained as a function of the
reheating temperature in the following way \cite{B4} 
\begin{equation}
N_{re}=\frac{1}{3(1+\omega )}\ln \left( \frac{3^{2}\cdot 5~V_{end}}{\gamma
\pi ^{2}\bar{g}_{\ast }T_{re}^{4}}\right) .  \label{Nr}
\end{equation}%
\begin{figure}[h!]
\includegraphics[width=12cm]{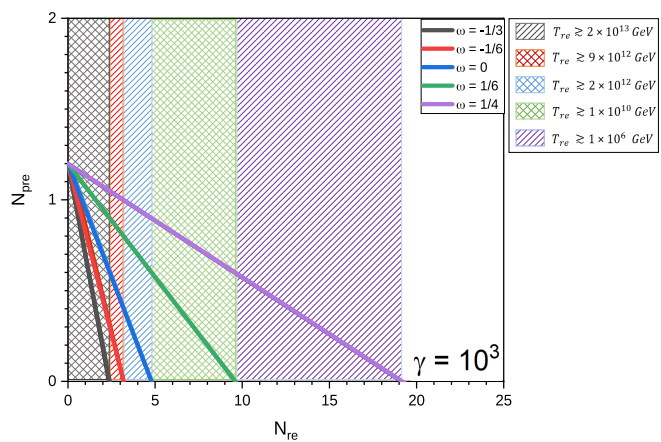}% Here is how to import EPS art
\caption{Variation of the preheating duration $N_{pre}$ as a function of
reheating e-folds for the case of $\protect\gamma =10^{3}$ considering
different values of the (EoS) parameter $\protect\omega $, each region is
determined by a reheating temperature bound, in the black dashed region the
temperature is bounded as $T_{re}\gtrsim 2\times 10^{13}GeV,$ the red region
is bouned as $T_{re}\gtrsim 9\times 10^{12}GeV,$ the blue region is bounded
by $T_{re}\gtrsim 2\times 10^{12}GeV,$ the green one is bounded by $%
T_{re}\gtrsim 1\times 10^{10}GeV,$ and the purple region is bounded by $%
T_{re}\gtrsim 1\times 10^{6}GeV$.}
\label{fig:1}
\end{figure}
\begin{figure}[h!]
\includegraphics[width=12cm]{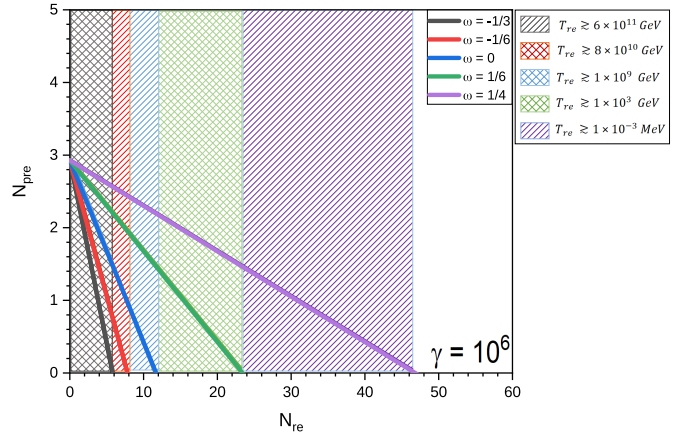}% Here is how to import EPS art
\caption{Variation of the preheating duration $N_{pre}$ as a function of
reheating e-folds for the case of $\protect\gamma =10^{6}$ considering
different values of the (EoS) parameter $\protect\omega $, each region is
determined by a reheating temperature bound, in the black dashed region the
temperature is bounded as $T_{re}\gtrsim 6\times 10^{11}GeV,$ the red region
is bouned as $T_{re}\gtrsim 8\times 10^{10}GeV,$ the blue region is bounded
by $T_{re}\gtrsim 1\times 10^{9}GeV,$ the green one is bounded by $%
T_{re}\gtrsim 1\times 10^{3}GeV,$ and the purple region is bounded by $%
T_{re}\gtrsim 1\times 10^{-3}MeV$.}
\label{fig:2}
\end{figure}
Figs. (\ref{fig:1},\ref{fig:2}) presents the variations of the preheating
e-folds number $N_{pre}$ as a function of the reheating duration $N_{re}$,
when reheating is instantaneous $N_{re}\rightarrow 0$\ the preheating
duration takes values independent on the (EoS) parameter, for the cases of $%
\gamma :$\ $10^{3},\ 10^{6}$\ all the lines with different values of $\omega 
$\ tends toward a preheating duration that lies between $\sim 1-3$\ e-folds.
Our conclusion is that models without preheating considered in previous
works estimate higher values of reheating duration, while here, considering
preheating as a phase would only conduct a rapid reheating transition step
toward the radiation-dominated era, for each value of $\omega $\ an instant
preheating duration is related to a minimum value of reheating temperature
which increases as we take lower values of (EoS), additionally, in order to
retain a successful Big Bang Nucleosynthesis (BBN), the thermalization is
required to be bounded as $T_{re}\gtrsim 4MeV$ \cite{C6,C9}, as a
consequence the lower bound on the reheating temperature for the case of $%
\omega =1/4$ with $\gamma =10^{6}$ must not be expected. Preheating e-folding%
$\ N_{pre}$ is linked to the inflationary quantities through $V_{end}$, $N$,
and $H_{k}$, these quantities were calculated for the E-model considered
previously. Furthermore, the preheating period is specified by a parameter $%
\gamma $ that relates the energy density at the end of inflation $\rho
_{end} $ to the preheating energy density $\rho _{pre}$. $N_{re}$\ can be
calculated considering the final reheating thermalization temperature. Here,
we must mention that in order to have a valid reheating duration $%
N_{re}\gtrsim 0$ we must consider the bound 
\begin{equation}
T_{re}\leq \left( \frac{3^{2}\cdot 5~V_{end}}{\gamma \pi ^{2}\bar{g}_{\ast }}%
\right) ^{\frac{1}{4}},  \label{bn}
\end{equation}%
this means that a reheating temperature is bounded as $T_{re}\lesssim
10^{13}GeV,$ the higher limit from the bound will cause an instant
transition of reheating which makes the preheating duration independent of
the choice of the (EoS) parameter. However, a lower temperature value can
increase the reheating duration, as a consequence $N_{pre}$ will directly
depend on the parameters $T_{re}$ and $\omega .$ Next, we will consider the
case of the two phases scenario where preheating and reheating occur
non-instantaneously. Our goal is to achieve new constraints on preheating
duration considering different cases of $T_{re}$\ and $\omega $\ which
characterize the reheating scenario.\ 

\subsubsection{\label{sec:221}Preheating and reheating as two phases scenario%
}

Inflation must end when the value of $\omega $ became larger than $-1/3$, in
order to satisfy the condition of energy density dominance and preserve the
causality $\omega $ must be smaller than $1$ \cite{A12}. Considering the
results shown in Fig. (\ref{fig:1},\ref{fig:2}) the choice of a specific
(EoS) value enables a reheating temperature bound, our purpose is to
constrain the preheating e-fold considering various cases of $T_{re}$ and $%
\omega $ provided by the previous findings. 
\begin{figure}[h!]
\includegraphics[width=9.5cm]{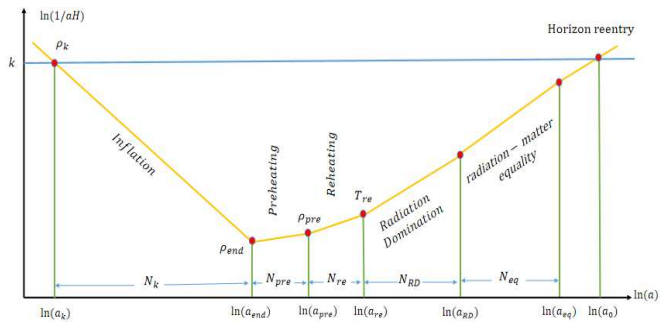}% Here is how to import EPS art
\caption{The evolution of the comoving Hubble scale $(aH)^{-1}$ connects the
inflationary phase with the CMB. The end of the inflation is denoted by $%
\protect\rho _{end}$, the end of preheating phase is denoted by $\protect%
\rho _{pre},$ and reheating is ended by the final temperature $T_{re}$. The
inflationary phase and radiation-dominated era connect through the
preheating and reheating phases.}
\label{fig:2_1}
\end{figure}
\begin{figure*}[tbp]
\includegraphics[width=18cm]{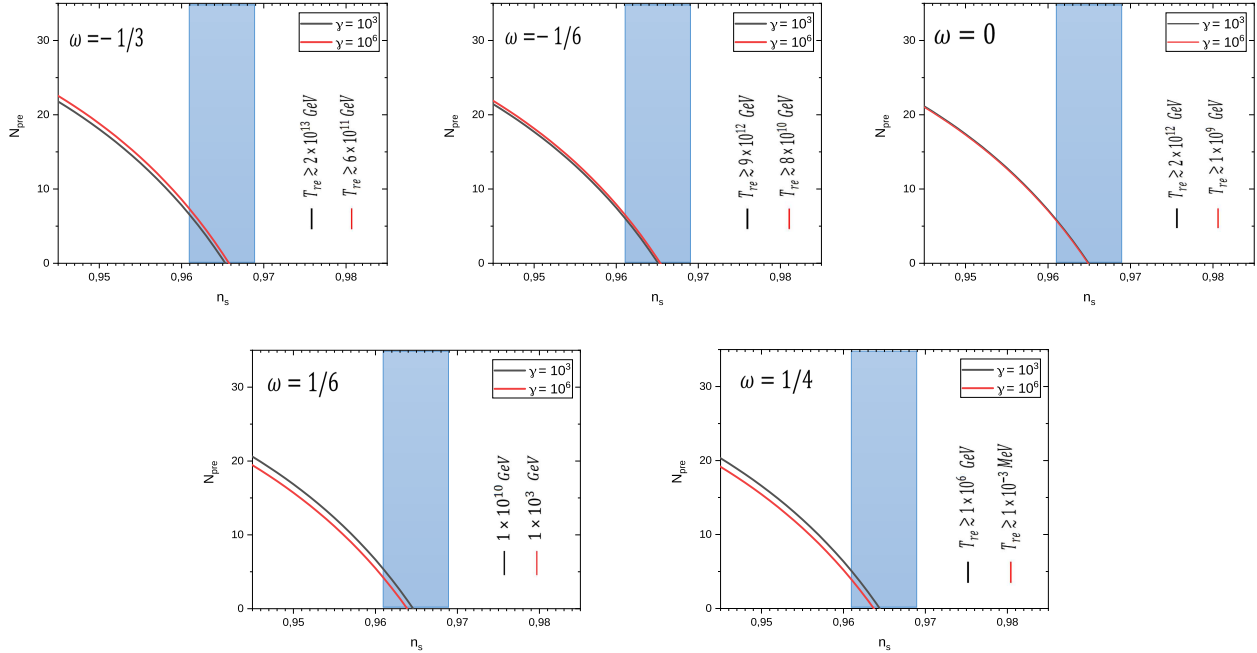}% Here is how to import EPS art
\caption{Constraints on the preheating duration $N_{pre}$ for the $\protect%
\alpha $-attractor E model potential considering both cases: $\protect\gamma %
=10^{3}$ and $\protect\gamma =10^{6}$ with different values of the
thermalization temperature and (EoS) parameter, we chose $n=1,\protect\alpha %
=0.1$. Here, the vertical light blue region represents Planck's bounds on $%
n_{s}=09649\pm 0.0042$ \protect\cite{A4}.}
\label{fig:2_2}
\end{figure*}
It is well known that the mechanism of nonperturbative effect which occurs
before reheating is what we call preheating. This process of parametric
resonance does not completely decay inflaton into the radiation field.
Therefore, a subsequent thermalization must take place and the perturbative
decay will be necessary to complete the reheating process. Our objective
here is to consider a two phases model to describe both preheating and
reheating durations reconcilably, Figure \ref{fig:2_1} describe our
methodology of previous calculations, the variation of preheating duration
as a function of the spectral index $n_{s}$ illustrated in Figure \ref%
{fig:2_2} show good consistency for a wide range of reheating temperature $%
T_{re}$ along with the (EoS) parameter. However, a lower bound on $T_{re}$
must be taken into consideration, knowing that according to \cite{C6} the
minimum reheating temperature must satisfy $T_{re}\geq 4MeV,$ we could
exclude the case $\omega =1/4$ with $\gamma =10^{6}$ for the two phases
scenario model. This analysis motivates us to subsequently analyze
preheating as a one-phase scenario with an instant transition of reheating,
and we will show that this model still holds.

\section{\label{sec:3}Primordial gravitational waves from preheating}

Because we are interested in the connection of gravity-wave energy density
spectrum with Planck data, we must convert the GW spectrum into actual
physical values. The present scale factor in comparison to the one when GW
production stops can be expressed as \cite{B5}

\begin{equation}
\Omega _{gw,0}h^{2}=\Omega _{gw}(f)~\left( \frac{\bar{g}_{\ast }}{\bar{g}_{0}%
}\right) ^{-1/3}\Omega _{r,0}h^{2}\left( \frac{a_{end}}{a_{pre}}\right)
^{4}~.
\end{equation}

The number of e-folds between the end of inflation to the time when
preheating completed can be written as $a_{end}/a_{pre}=e^{-N_{pre}}$, which
makes the current gravity-wave energy density spectrum related to the
preheating e-folds considering $\omega =1/3$ given as 
\begin{eqnarray}
\Omega _{gw,0}h^{2} &=&\Omega _{gw}(f)\left( \frac{\bar{g}_{\ast }}{\bar{g}%
_{0}}\right) ^{-1/3} \\
&&\Omega _{r,0}h^{2}\exp \left( -4\left[ 61.6-\frac{1}{4}\ln \left( \frac{%
V_{end}}{\gamma H_{k}^{4}}\right) -N\right] \right) .  \nonumber
\end{eqnarray}%
\begin{figure}[h!]
\includegraphics[width=12cm]{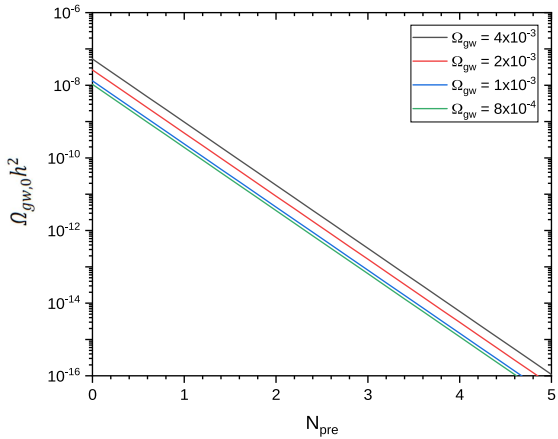}% Here is how to import EPS art
\caption{The variation of the density spectra of GWs as a function of $%
N_{pre}$. For different values of gravity wave energy density $\Omega _{gw}$%
, the grey region corresponds to the bound $\Omega _{gw,0}h^{2}\leq 1.85$\ $%
\times 10^{-6}$ \protect\cite{A4}.}
\label{fig:3}
\end{figure}
From Fig. \ref{fig:3} the variation of $\Omega _{gw,0}h^{2}$ \ as a function
of $N_{pre}$\ for some fixed values of $\Omega _{gw}$\ are presented, we
plotted the energy density spectrum\ $\Omega _{gw,0}h^{2}$\ as a function of
the preheating duration $N_{pre},$\ taking the present GW spectra to be\ $%
8\times 10^{-4}\leq \Omega _{gw}\leq 4$\ $\times 10^{-3}$, when $%
N_{pre}\rightarrow 0$\ the GW density spectrum takes an initial value for
all the cases with different $\Omega _{gw}.$ When we increase the present GW
energy density, the initial values that correspond to $\Omega
_{gw,0}h^{2}(N_{pre}=0)$ increases as well, knowing that values of
preheating with GW density spectrum bounded as \ $\Omega _{gw,0}h^{2}\leq
1.85$\ $\times 10^{-6}$ are most compatible according to Planck's results 
\cite{A4,B1}. 
\begin{figure}[h!]
\includegraphics[width=12cm]{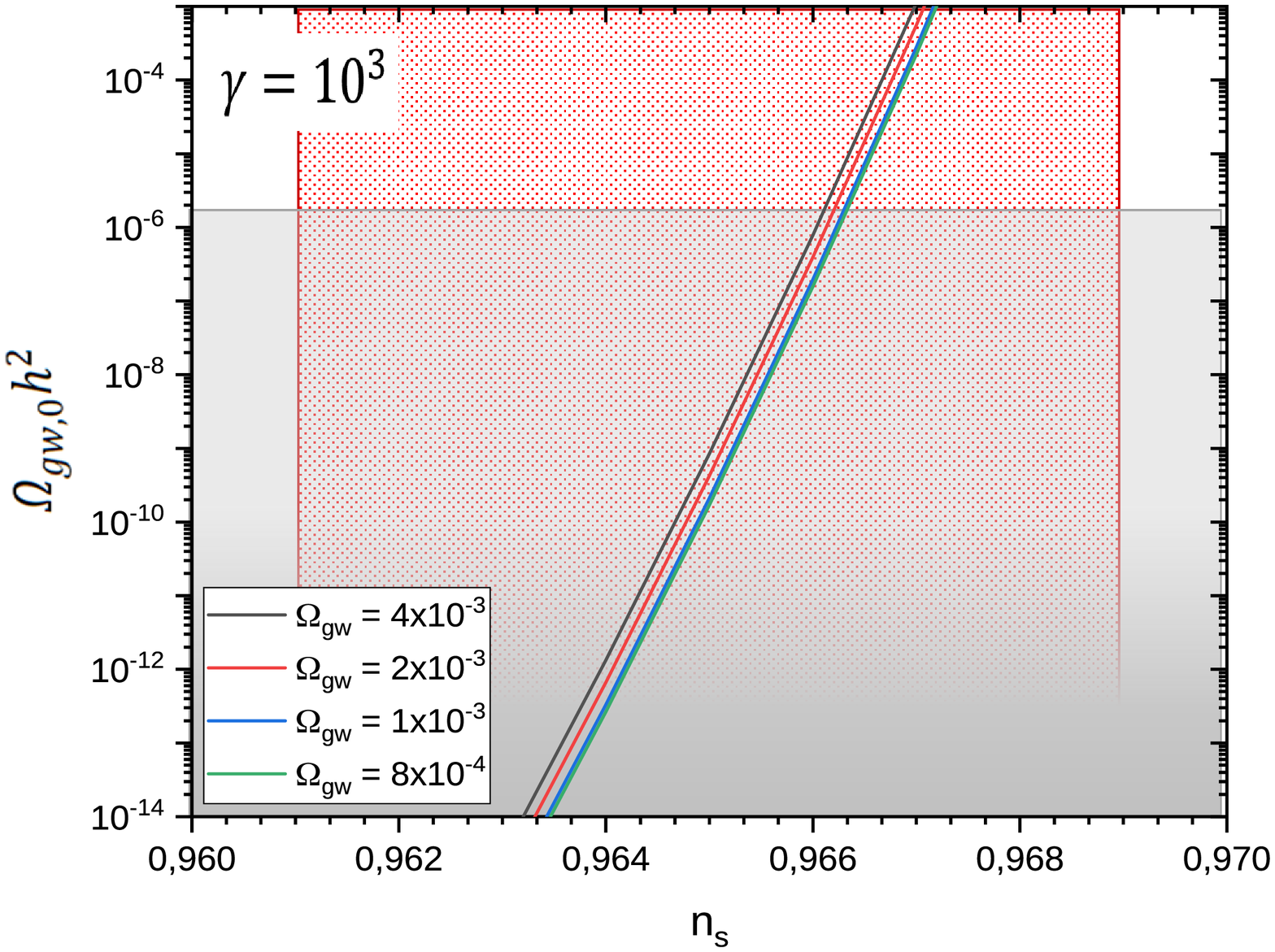}% Here is how to import EPS art
\caption{The variation of the density spectra of GWs as a function of the
spectral index $n_{s}$. For different values of gravity wave energy density $%
\Omega _{gw}$, we choose $8\cdot 10^{-4}\leq \Omega _{gw}\leq 4\cdot 10^{-3}$%
, and $\protect\gamma =10^{3}$. The black line corresponds to $\Omega
_{gw}=4\cdot 10^{-3}$, red line corresponds to $\Omega _{gw}=2\cdot 10^{-3}$%
, blue line corresponds to $\Omega _{gw}=10^{-3}$, green line corresponds to $%
\Omega _{gw}=8\cdot 10^{-4}$,the red shaded region represents Planck's
bounds on $n_{s}$ and the grey region corresponds to the bound $\Omega
_{gw,0}h^{2}\leq 1.85$\ $\times 10^{-6}$ \protect\cite{A4}.}
\label{fig:4}
\end{figure}
\begin{figure}[h!]
\includegraphics[width=12cm]{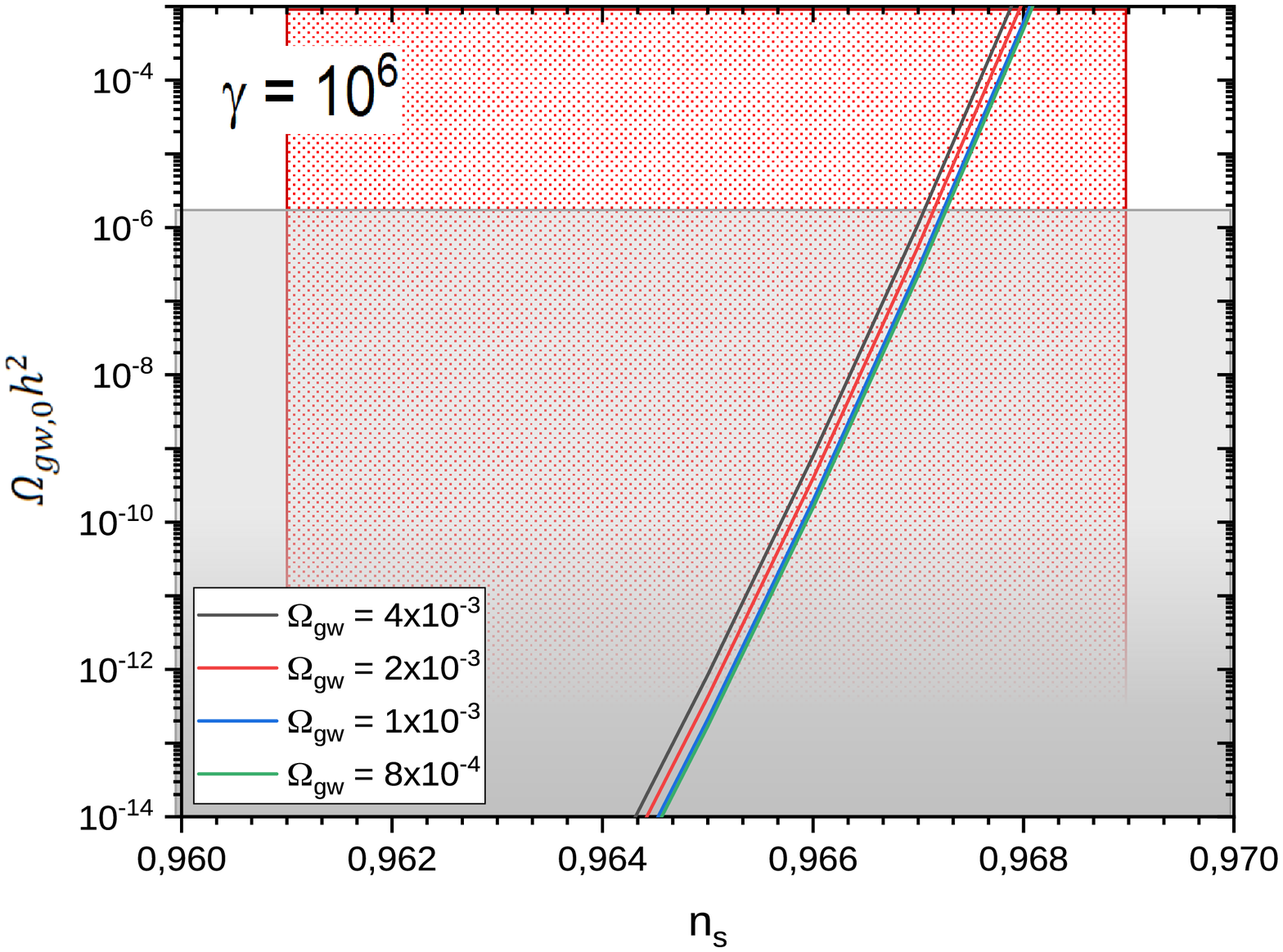}% Here is how to import EPS art
\caption{The variation of the density spectra of GWs as a function of the
spectral index $n_{s}$. For different values of gravity wave energy density $%
\Omega _{gw}$, we choose $8\cdot 10^{-4}\leq \Omega _{gw}\leq 4\cdot 10^{-3}$%
, and $\protect\gamma =10^{6}$. The black line corresponds to $\Omega
_{gw}=4\cdot 10^{-3}$, red line corresponds to $\Omega _{gw}=2\cdot 10^{-3}$%
, blue line corresponds to $\Omega _{gw}=10^{-3}$, green line corresponds to $%
\Omega _{gw}=8\cdot 10^{-4}$, the red shaded region represents Planck's
bounds on $n_{s}$ and the grey region corresponds to the bound $\Omega
_{gw,0}h^{2}\leq 1.85$\ $\times 10^{-6}$ \protect\cite{A4}.}
\label{fig:5}
\end{figure}
Figures \ref{fig:4} and \ref{fig:5} demonstrate the evolution of the GW
density spectrum with respect to the spectral index $n_{s}$ for different
values of the current GW spectra $\Omega _{gw}$, we plot $\Omega _{gw,0}h^{2}
$ considering the expansion of the Universe from the end of inflation up to
later times of preheating, because of the bound of the final temperature of
reheating $T_{re}\lesssim 10^{13}GeV,$ the duration $N_{re}$ from Eq. (\ref%
{Nr}) could be considered as instantaneous, which make preheating duration
minimally dependent on the (EoS) parameter $\omega $ as observed in Figs. (%
\ref{fig:1},\ref{fig:2}). We choose the values of $\gamma :$\ $10^{3},\
10^{6}$ from the previous analysis\ which presents good consistency$\ $%
according to Planck's results, the curves decrease as the GW energy density
became very negligible $\Omega _{gw}\rightarrow 0$. For both cases $\gamma
=10^{3},$ and $\gamma =10^{6}$\ with $\Omega _{gw,0}h^{2}\leq 1.85$\ $\times
10^{-6}$ all the lines with different $\Omega _{gw}$\ tends towards the
central value of \ the spectral index $n_{s}$.

\section{\label{sec:4}Conclusion}

After we examine the basic equations that describe gravity waves production
using metric perturbation and calculate $\Omega _{gw,0}h^{2}$ that
represents the abundance of gravity wave energy density today, we discuss
the E model $\alpha $-attractor and calculate the expression of the
observational parameters $n_{s}$ and $r$, we compute these parameters as
functions of inflation's e-folds $N$ and test the compatibility of the
parameter $\alpha $ with Planck's results,\ our results show that small
values of $\alpha $ give good results on the tensor-to-scalar ratio
parameter, while the spectral index $n_{s}$\ is best estimated for $N=60$.
We derive the preheating duration as functions of inflationary parameters in
Eq. (\ref{pr}), and consider different thermalization temperature to discuss
the case when the thermalization temperature is low enough to conduct to two
phases scenarios where preheating and reheating occurs for specific
durations. we provide additional constraints that estimate the bound on the
reheating temperature as $T_{re}\lesssim 10^{13}GeV$ \ in order to have a
valid reheating duration $N_{re}\gtrsim 0.$ Our results suggest that when $%
N_{re}\rightarrow 0$ the preheating e-folds is independent on the choice of
the (EoS), the duration of preheating is plotted as a function of the
spectral index for the two models we considered previously, we show that $%
N_{pre}$ is weakly sensitive to $\gamma $ and independent on the E model
inflation parameter $n$. We finally calculate the current gravity wave
energy density spectrum as a function of the duration $N_{pre}$, which is a
possible alternative\ to study GW density spectrum according to recent
Planck's results. Assuming the density parameter $\Omega _{gw}$ to be
bounded as $8\times 10^{-4}\leq \Omega _{gw}\leq 4$\ $\times 10^{-3}$, we
chose $n=1,\alpha =0.1$, and $\gamma :$\ $10^{3},\ 10^{6}$, and found that
both cases where $\gamma =10^{3},\gamma =10^{6}$ show good consistency with
observation. We conclude that the E model $\alpha $-attractor parameters
appear to have consistent results considering the preheating duration,
regardless of whether the process is instant or takes a certain number of
e-folds to complete, once we determine the final temperature of
thermalization $T_{re}$, other preheating parameters are determined using
the E model $\alpha $-attractor inflation. As a result, it would be
interesting to investigate the physics of preheating in the context of
Primordial GWs.


\begin{thebibliography}{99}
\bibitem{A1} Linde, A. (2008). Inflationary cosmology. In Inflationary
Cosmology (pp. 1-54). Springer, Berlin, Heidelberg.

\bibitem{A2} Gorbunov, D. S., \& Rubakov, V. A. (2011). Introduction to the
theory of the early universe: Cosmological perturbations and inflationary
theory. World Scientific.

\bibitem{A3} Lyth, D. H., \& Riotto, A. (1999). Particle physics models of
inflation and the cosmological density perturbation. Physics Reports,
314(1-2), 1-146.

\bibitem{A4} Collaboration, P., Aghanim, N., Akrami, Y., Ashdown, M.,
Aumont, J., Baccigalupi, C., ... \& Rosset, C. (2020). Planck 2018 results.
VI. Cosmological parameters.

\bibitem{A5} Kallosh, R., \& Linde, A. (2013). Superconformal
generalizations of the Starobinsky model. Journal of Cosmology and
Astroparticle Physics, 2013(06), 028.

\bibitem{A6} Scalisi, M. (2015). Cosmological $\alpha $-attractors and de
Sitter landscape. Journal of High Energy Physics, 2015(12), 1-15.

\bibitem{A7} Abbott, L. F., Farhi, E., \& Wise, M. B. (1982). Particle
production in the new inflationary cosmology. Physics Letters B, 117(1-2),
29-33.

\bibitem{A8} Dolgov, A. D., \& Linde, A. D. (1982). Baryon asymmetry in the
inflationary universe. Physics Letters B, 116(5), 329-334.

\bibitem{A9} Ueno, Y., \& Yamamoto, K. (2016). Constraints on $\alpha $%
-attractor inflation and reheating. Physical Review D, 93(8), 083524.

\bibitem{A10} Shtanov, Y., Traschen, J., \& Brandenberger, R. (1995).
Universe reheating after inflation. Physical Review D, 51(10), 5438.

\bibitem{A11} Cook, J. L., Dimastrogiovanni, E., Easson, D. A., \& Krauss,
L. M. (2015). Reheating predictions in single field inflation. Journal of
Cosmology and Astroparticle Physics, 2015(04), 047.

\bibitem{A12} El Bourakadi, K., Bousder, M., Sakhi, Z., \& Bennai, M.
(2021). Preheating and reheating constraints in supersymmetric braneworld
inflation. The European Physical Journal Plus, 136(8), 1-19.

\bibitem{A13} Sakhi, Z., et al. "Effect of brane tension on reheating
parameters in small field inflation according to Planck-2018 data."
International Journal of Modern Physics A 35.30 (2020): 2050191.

\bibitem{A14} Traschen, J. H., \& Brandenberger, R. H. (1990). Particle
production during out-of-equilibrium phase transitions. Physical Review D,
42(8), 2491.

\bibitem{A15} Shtanov, Y., Traschen, J., \& Brandenberger, R. (1995).
Universe reheating after inflation. Physical Review D, 51(10), 5438.

\bibitem{A16} Fu, C., Wu, P., \& Yu, H. (2017). Inflationary dynamics and
preheating of the nonminimally coupled inflaton field in the metric and
Palatini formalisms. Physical Review D, 96(10), 103542.

\bibitem{A17} Fu, C., Wu, P., \& Yu, H. (2019). Nonlinear preheating with
nonminimally coupled scalar fields in the Starobinsky model. Physical Review
D, 99(12), 123526.

\bibitem{A18} Khlebnikov, S., \& Tkachev, I. (1997). Relic gravitational
waves produced after preheating. Physical Review D, 56(2), 653.

\bibitem{A19} Garcia-Bellido, J., \& Figueroa, D. G. (2007). Stochastic
background of gravitational waves from hybrid preheating. Physical review
letters, 98(6), 061302.

\bibitem{A20} Dufaux, J. F., Bergman, A., Felder, G., Kofman, L., \& Uzan,
J. P. (2007). Theory and numerics of gravitational waves from preheating
after inflation. Physical Review D, 76(12), 123517.

\bibitem{A21} Carrasco, J. J. M., Kallosh, R., \& Linde, A. (2015). $\alpha $%
-attractors: Planck, LHC and dark energy. Journal of High Energy Physics,
2015(10), 1-20.

\bibitem{A22} Carrasco, J. J. M., Kallosh, R., \& Linde, A. (2015).
Cosmological attractors and initial conditions for inflation. Physical
Review D, 92(6), 063519.

\bibitem{A23} Mishra, S. S., Sahni, V., \& Starobinsky, A. A. (2021). Curing
inflationary degeneracies using reheating predictions and relic
gravitational waves. Journal of Cosmology and Astroparticle Physics,
2021(05), 075.

\bibitem{B1} Li, J., Yu, H., \& Wu, P. (2020). Production of gravitational
waves during preheating in $\alpha $-attractor inflation. Physical Review D,
102(8), 083522.

\bibitem{B2} Adshead, P., Giblin Jr, J. T., \& Weiner, Z. J. (2018).
Gravitational waves from gauge preheating. Physical Review D, 98(4), 043525.

\bibitem{B3} Dufaux, J. F., Bergman, A., Felder, G., Kofman, L., \& Uzan, J.
P. (2007). Theory and numerics of gravitational waves from preheating after
inflation. Physical Review D, 76(12), 123517.

\bibitem{B4} El Bourakadi, K., Ferricha-Alami, M., Filali, H., Sakhi, Z., \&
Bennai, M. (2021). Gravitational waves from preheating in Gauss--Bonnet
inflation. The European Physical Journal C, 81(12), 1-8.

\bibitem{B5} Podolsky, D., Felder, G. N., Kofman, L., \& Peloso, M. (2006).
Equation of state and beginning of thermalization after preheating. Physical
Review D, 73(2), 023501.

\bibitem{B6} Dufaux, J. F., Felder, G. N., Kofman, L., Peloso, M., \&
Podolsky, D. (2006). Preheating with trilinear interactions: Tachyonic
resonance. Journal of Cosmology and Astroparticle Physics, 2006(07), 006.

\bibitem{C1} Kallosh, R., Linde, A., \& Roest, D. (2013). Superconformal
inflationary $\alpha $-attractors. Journal of High Energy Physics, 2013(11),
1-13.

\bibitem{C2} Dai, L., Kamionkowski, M., \& Wang, J. (2014). Reheating
constraints to inflationary models. Physical review letters, 113(4), 041302.

\bibitem{C3} Drewes, M., Kang, J. U., \& Mun, U. R. (2017). CMB constraints
on the inflaton couplings and reheating temperature in $\alpha $-attractor
inflation. Journal of High Energy Physics, 2017(11), 1-42.

\bibitem{C4} Dai, L., Kamionkowski, M., \& Wang, J. (2014). Reheating
constraints to inflationary models. Physical review letters, 113(4), 041302.

\bibitem{C5} Lozanov, K. D., \& Amin, M. A. (2017). Equation of state and
duration to radiation domination after inflation. Physical review letters,
119(6), 061301.

\bibitem{C6} Hannestad, S. (2004). What is the lowest possible reheating
temperature?. Physical Review D, 70(4), 043506.

\bibitem{C7} Haque, M. R., Maity, D., \& Saha, P. (2020). Two-phase
reheating: CMB constraints on inflation and dark matter phenomenology.
Physical Review D, 102(8), 083534.

\bibitem{C8} Salamate, F., Khay, I., Ferricha-Alami, M., Chakir, H., \&
Bennai, M. (2019). E-Model $\alpha $-Attractor on Brane from Planck Data and
Reheating Temperature. Moscow University Physics Bulletin, 74(5), 537-543.

\bibitem{C9} Kawasaki, M., Kohri, K., \& Sugiyama, N. (2000). MeV-scale
reheating temperature and thermalization of the neutrino background.
Physical Review D, 62(2), 023506.

\bibitem{D1} Li, J., Yu, H., \& Wu, P. (2020). Production of gravitational
waves during preheating in $\alpha $-attractor inflation. Physical Review D,
102(8), 083522.\~{A}~
\end{thebibliography}
\end{document}